\documentstyle[tighten,preprint,aps]{revtex}
\begin{document}
\title{ Exclusion statistics and many-particle states}
\author{P. Mitra\thanks{e-mail mitra@saha.ernet.in}}
\address{Saha Institute of Nuclear Physics,
Block AF, Bidhannagar,
Calcutta 700 064, INDIA}
\author{and}
\address{Scuola Internazionale Superiore di Studi Avanzati,
Via Beirut 2-4,
34014 Trieste, ITALY}
\date{November 1994\\SISSA-191/94/EP\\hep-th/9411236}
\maketitle
\begin{abstract}
The thermodynamic distribution function for exclusion
statistics is derived. Creation and
annihilation operators for particles obeying such statistics
are discussed. A connection with anyons is pointed out.
\end{abstract}

Exclusion statistics, which is a generalized version of Fermi- Dirac
statistics, was introduced a few years back \cite{Hald}. While
physical particles in the three dimensional world obey either
Bose-Einstein or Fermi-Dirac statistics, this generalized
statistics is expected to find application in  regions of
condensed matter where the number of spatial dimensions is in
effect reduced. There has been an enormous amount of
interest in the subject in the past one year \cite{Shank,Wu,Ha,Wil,Nair},
no doubt because of the wide range of applications envisaged.

It was mentioned in \cite{Hald} that no microscopic theory was introduced
there:  a phenomenological framework for appropriately detailed
theories was presented. However, the basic definition of exclusion statistics,
whereby the exclusion of particles from single-particle
states is in general less limited than in Fermi statistics,
tempts one to follow the standard discussion
of fermionic thermodynamics with the minor changes that seem to be
necessary. Statistical thermodynamics has recently
been generalized in this spirit \cite{Wu}.
The distributions thus obtained have  been made use of in \cite{Wil,Nair}.

Although the starting point of \cite{Wu} for the number of ways of
putting particles in a collection of single-particle states
represents an interpolation
between the  forms which are known to be valid
for Bose and Fermi statistics, it is not supported by a direct
combinatorial argument, {\it except in the limit of large collections
of single-particle states} \cite{Wil}.  However
very large collections or {\it cells} cannot be considered together unless
the temperature is very high, so departures from the distribution
found in this way can be expected. Furthermore, one runs into negative
probabilities even for simple, rational values of the degree of exclusion
\cite{Wil}. It therefore becomes necessary
to derive the generalized distribution without appealing
to large cells. That is the purpose of this letter.

As mentioned above, the distribution found in \cite{Wu} has been
used in \cite{Nair}, where the algebra of creation and annihilation
operators for oscillators obeying this distribution has been derived.
In the same way we shall  discuss creation and annihilation of oscillators
obeying the distribution derived in this letter, albeit in a much more
limited way than \cite{Nair}. An interesting connection with anyons will
emerge.

In the original formulation \cite{Hald}, a degree $g$ of exclusion
may be said to exist if $n$ particles in a single-particle
state reduce the dimensionality of the Hilbert space available to an
additional particle by a fraction $gn$. We
have omitted here the indices introduced in \cite{Hald} to accommodate
several species. This degree is clearly 0 and 1 for the normal bosonic
and fermionic cases respectively. In general $g$ has to be
rational, but it is simplest when ${1\over g}$ is a positive integer, $m$
say. Then one may define the exclusion condition by stating that
a given single-particle state may accommodate only a maximum of $m$ particles.
This is the definition which will be used hereafter.

The approach of \cite{Wu} is to approximate the number of independent
ways of putting $n$ particles in $d$ single-particle states  by
\begin{equation}
W={[d+(1-{1\over m})(n-1)]!\over n![d-1-{1\over m}(n-1)]!}.\end{equation}
This expression is exact for the bosonic and fermionic limits.
One may seek to replace it for the general case
by the number obtained directly from the
definition of exclusion given above. But there is a simpler procedure
avoiding combinatorics.

The grand canonical partition function of a system of particles with
chemical potential $\mu$ at a temperature $\beta^{-1}$ is
\begin{equation}
Z=\sum e^{-\beta(E-\mu N)},\end{equation}
where the sum is over different states {\it of the
system as a whole} with energies $E$ and numbers
$N$ of particles. If the energy can be assumed to be additive, one can
rewrite this as
\begin{equation}
Z=\sum e^{-\beta\sum n_i(\epsilon_i-\mu)},\end{equation}
where the grand sum is now over the occupation numbers $n_i$ of the
different single-particle states with respective energies $\epsilon_i$.
A state of the system is supposed to be uniquely
specified by the occupation numbers of the individual
single-particle states. The grand sum factorizes,
\begin{equation}
Z=\prod_i\sum_{n_i} e^{-\beta n_i(\epsilon_i-\mu)},\end{equation}
each factor corresponding to one particular single-particle state.
Upto this stage everything is standard. The sum over $n_i$ is an
infinite geometric series for bosons because all nonnegative integral
values are allowed. For fermions, the sum has only two terms, because
the occupation number must be either zero or unity. In the case of
particles obeying exclusion statistics, the allowed values of $n_i$
are 0, 1, 2, ... $m$. The finite geometric sum
over $n_i$ can again be found:
\begin{equation}
Z=\prod_i {e^{-\beta(m+1)(\epsilon_i-\mu)}-1\over
e^{-\beta(\epsilon_i-\mu)}-1}.\end{equation}
This grand partition function can be used to calculate all
thermodynamic quantities. In particular, it is of interest to
write down the average number of particles:
\begin{equation}
\bar N=\sum_i [ (e^{\beta(\epsilon_i-\mu)}-1)^{-1}-(m+1)
(e^{\beta(m+1)(\epsilon_i-\mu)}-1)^{-1}].\end{equation}
As usual, this is interpreted to mean that the average occupation
number of the $i$-th single-particle state is
\begin{equation}
\bar{n_i}= [ (e^{\beta(\epsilon_i-\mu)}-1)^{-1}-(m+1)
(e^{\beta(m+1)(\epsilon_i-\mu)}-1)^{-1}].\end{equation}
This can equivalently be written as
\begin{equation}
\bar{n_i}={e^{(m-1)\beta(\epsilon_i-\mu)}+...+
(m-1)e^{\beta(\epsilon_i-\mu)}+m\over e^{m\beta(\epsilon_i-\mu)}
+...+e^{\beta(\epsilon_i-\mu)}+1}.\label{n}\end{equation}
This generalizes the Fermi form in a straightforward way.
At zero temperature, all single-particle states with energies
below $\mu$ are fully occupied {\it i.e.},
have an occupation number exactly equal to $m$ and all other states are empty,
as in the distribution function of \cite{Wu}. However, at finite
temperatures, the functions are different. Note that the right hand side
of (\ref{n}) lies between $m$ and 0 and decreases monotonically
as $e^{\beta(\epsilon_i-\mu)}$ increases.

It is instructive to consider the simplest case of new statistics,
{\it viz.}, $m=2$. Here the average occupation number is
\begin{equation}
\bar{n_i}={e^{\beta(\epsilon_i-\mu)}+2\over e^{2\beta(\epsilon_i-\mu)}
+e^{\beta(\epsilon_i-\mu)}+1}.\end{equation}
This equation can be inverted by defining
\begin{equation}
x=e^{\beta(\epsilon-\mu)}\end{equation}
and solving a quadratic in $x$ in terms of $\bar n$:
\begin{equation}
x={\sqrt{1+6\bar n-3\bar n^2}+1-\bar n\over 2\bar n}.\label{d}\end{equation}
This root is always real and positive because $\bar n$ cannot exceed 2.

The method of \cite{Nair} can be used to find the algebra of creation
and annihilation operators corresponding to the distribution (\ref{d}):
\begin{equation}
aa^\dagger={\sqrt{1+6N-3N^2}+1-N\over 2N}a^\dagger a.\label{al}\end{equation}
Here the standard notation for oscillator operators is used, so that
$N$ is now the number operator for the oscillator.
This relation looks complicated, but it is easy to see that
it corresponds to the matrix representation
\begin{equation}
a=\pmatrix{0 & 1 & 0\cr 0&0&1\cr 0&0&0}.\end{equation}
This is different from that found in \cite{Nair} on the basis of the
distribution of \cite{Wu}. The number operator, which has the matrix
representation
\begin{equation}
N=\pmatrix{0 & 0 & 0\cr 0&1&0\cr 0&0&2},\end{equation}
can be expressed as
\begin{equation}
N=a^\dagger a-aa^\dagger +1.\end{equation}
Although the operator algebra (\ref{al}) is satisfied, the matrices
also satisfy the much simpler q-oscillator algebra
\begin{equation}
aa^\dagger -qa^\dagger a=q^{-N},\end{equation}
with $q=\exp (-{i\pi\over 3})$. Thus the particles satisfying
exclusion statistics of degree 1/2 appear to be associated with the
q-value of not $\pm i$ as might have been na\"\i vely expected, but
$\exp (-{i\pi\over 3})$.

q-commutation enters the discussion in another way if one
considers different states of the system. The allowed range of occupation
numbers of a single-particle state have been taken to be unaffected
by the occupation numbers of the other states. This corresponds to
product Hilbert spaces and
essentially independent creation/annihilation operators for different
single-particle states. One can write
\begin{equation}
{a_i}^\dagger {a_j}^\dagger =q_{ij}{a_j}^\dagger {a_i}^\dagger,\end{equation}
where $q_{ij}$ is a new phase factor. The
phase factor for $i\ne j$ is in general
a matter of convention, but it turns out that a very natural choice
is \cite{Nair}
\begin{equation}
q_{ij}=\exp({2\pi i\over 3}\epsilon_{ij}).\label{nat}\end{equation}
To see this, note that the q-{\it symmetrized} wavefunction of a three-particle
state can be constructed out of three single-particle wavefunctions
$f^A_i$ (with $A$=1,2,3) \cite{PM} as
\begin{equation}
{1\over 3!}[f^1_if^2_jf^3_k + q_{ij}^{-1}f^1_jf^2_if^3_k
+q_{jk}^{-1}f^1_if^2_kf^3_j+q_{ik}^{-1}q_{jk}^{-1}f^1_kf^2_if^3_j
+q_{ij}^{-1}q_{ik}^{-1}f^1_jf^2_kf^3_i+
q_{ij}^{-1}q_{ik}^{-1}q_{jk}^{-1}f^1_kf^2_jf^3_i].
\end{equation}
This $(-q)-determinant$ is in general nonvanishing, but if the three
wavefunctions $f^A$ are identical, it vanishes with the  choice
(\ref{nat}) because of the property
\begin{equation}
1+q+q^2=0.\end{equation}
The generalized exclusion principle with $m=2$ is
then extended to states obtained by superposing the original
states labelled by the indices $i$.

We have considered the simplest case $m=2$ in detail.
For higher values of $m$, the expression for
$\bar n$ contains powers of $x$ higher than 2 and more complicated
equations have to be solved to obtain $x$ and hence the algebras. We
shall not do that, but must point out that (\ref{nat}) generalizes to
\begin{equation}
q_{ij}=\exp({2\pi i\over m+1}\epsilon_{ij}).\end{equation}
The $(-q)-determinant$ form \cite{PM} of the $(m+1)$-particle wavefunction
vanishes again if all the single-particle wavefunctions are the
same. This provides more direct evidence than \cite{Shank} that
anyons with these q-values obey exclusion statistics.

To conclude, we have derived the thermodynamic distribution function
for analogues of fermions where a single-particle state can accommodate
a limited number of particles. This is expected to find applications in
condensed matter physics.

\acknowledgments
I wish to thank SISSA for its hospitality.

\end{document}